# DESIGN ARBITRARY SHAPED 2D ACOUSTIC CLOAK WITHOUT SINGULARITY

**Jin Hu**
School of Aerospace Engineering, Beijing Institute of Technology
Beijing, P. R. China

**Xiaoming Zhou**
School of Aerospace Engineering, Beijing Institute of Technology
Beijing, P. R. China

**Gengkai Hu**[*]
School of Aerospace Engineering, Beijing Institute of Technology
Beijing, P. R. China

## ABSTRACT

A method is proposed to design arbitrary shaped two dimensional (2D) isotropic-inertia acoustic cloaks without singularity. The method is based on the deformation view of the transformation method, where the transformation tensor **A** is identified as the deformation gradient tensor and the transformed material parameters can be expressed by the principal stretches in the principal system of the deformation. The infinite material parameters of a perfect 2D cloak is induced by an infinite principal stretch in one direction while the other two remains finite at the inner boundary during the transformation. To circumvent this difficulty, for a 2D cloak we can choose the principal stretch perpendicular to the cloak plane to be also infinite but in the same order as the infinite principal stretch in the cloak plane during the transformation, so the transformed material parameters may keep finite. To illustrate this idea, the analytical expressions of nonsingular material parameters for a cylindrical acoustic cloak are given. For the acoustic cloaks with irregular shapes, the numerical method is proposed to evaluate the principal stretches and in turn the nonsingular material parameters. The designed 2D cloaks are validated by numerical simulation.

## INTRODUCTION

There has recently been a growing interest in use of coordinate transformation method [1-3] to design devices that control electromagnetic fields in prescribed ways. The most interesting application of this method might be the invisible cloak [2,4], which can make itself and the cloaked region transparent to electromagnetic waves. Parallel to the electromagnetic case, with acoustic metamaterials [5], acoustic cloaks can also be designed with the coordinate transformation method. The material parameters for realizing a two dimensional (2D) cylindrical acoustic cloak is proposed by Cummer et al. [6] based on 2D acoustic equations. Chen et al. [7] derive the general expressions of transformed acoustic material parameters by using the invariant property of the acoustic equation for scalar pressure under transformation, and the derived parameters for a three dimensional (3D) spherical cloak are the same as that proposed by Cummer et al. [8] by scattering theory. All the proposed ideal acoustic cloak implies an infinite mass at the inner boundary of the cloak. To remedy this, Norris [9] suggests an imperfect acoustic cloak which inner boundary is expended from a finite area instead of a point. Chen et al. [10] propose a reduced 2D cylindrical acoustic cloak with finite mass by keeping its principal refractive indexes unchanged, which is similar to the case of the reduced electromagnetic cloak [4]. Norris [10] also shows that for a given transformation the material composition of an acoustic cloak is not uniquely defined, thus a perfect isotropic-inertia acoustic cloak can be constructed with finite mass. Unfortunately, one component of the anisotropic modulus of such isotropic-inertia acoustic cloak becomes infinite at the inner boundary; this singularity can also limit the application of the acoustic cloak. The nonsingular acoustic cloak is important for real engineering applications, however, as far as we know there is no general method that can design nonsingular 2D acoustic cloak with arbitrary shape. Recently, Hu et al. [11] propose a method to remove the singularity of an arbitrary 2D electromagnetic cloak completely within the frame of coordinate transformations, so the theory can be extended to acoustic case. In this paper, we will employ this idea to design nonsingular acoustic cloak with arbitrary shape. The paper will be arranged as follows: the design method will be explained in Section 2 with help of design for a nonsingular cylindrical acoustic cloak; the method to design a nonsingular arbitrary

---

[*] *Corresponding author: hugeng@bit.edu.cn*

shaped acoustic cloak will be presented in Section 3, followed by discussion and conclusion.

## DESIGN METHOD OF NONSINGULAR ACOUSTIC CLOAK

The proposed method for nonsingular acoustic cloak is based on the deformation view of the acoustic transformation method recently established by Hu et al. [12]. The transformation material parameters are given by [7]

$$\boldsymbol{\rho}'^{-1} = \mathbf{A}\boldsymbol{\rho}^{-1}\mathbf{A}^T / \det \mathbf{A}, \quad \kappa' = \kappa \det \mathbf{A}. \quad (1)$$

where $\mathbf{A}$ is the Jacobian transformation tensor with components $A_{ij} = \partial x'_i / \partial x_j$, characterizing the mapping from the original space $\Omega$ to the transformed space $\Omega'$. Since $\mathbf{A}$ can be considered as the deformation gradient tensor for this spatial transformation, Hu et al [12] have shown that if the material parameters in the original space are homogeneous and isotropic (for the simplicity, we let them equal to 1), then the transformed material parameters can be expressed in a clear way as [12]

$$\boldsymbol{\rho}' = \text{diag}[\frac{\lambda_2\lambda_3}{\lambda_1}, \frac{\lambda_3\lambda_1}{\lambda_2}, \frac{\lambda_1\lambda_2}{\lambda_3}], \kappa' = \lambda_1\lambda_2\lambda_3. \quad (2)$$

where $\lambda_i$ ($i$=1,2,3) are the principal stretches of the deformation and the anisotropic mass density $\boldsymbol{\rho}'$ is expressed in the principal system of the deformation. In order to make a perfect cloak, its outer boundary needs to be fixed, i.e. $\mathbf{x}' = \mathbf{x}$ [9]. This condition naturally constraints the stretches to unity (without deformation) in the tangential directions at the outer boundary.

Equation (2) gives a clear explanation on how the singularity is formed during the construction of an acoustic cloak. Since the inner boundary of a perfect acoustic cloak is enlarged by a point, one (for 2D case) or two (for 3D case) of the principal stretches will become infinite there. Taking a 2D cylindrical cloak and a 3D spherical cloak as examples, the linear transformation of a cylindrical cloak is $r' = a + \frac{b-a}{b}r$, $\theta' = \theta$, $z' = z$ and the cloak is bordered by $r' \in (a,b)$, the principal stretches of each point within the cloak are given by

$$\lambda_r = \frac{dr'}{dr} = \frac{b-a}{b},$$
$$\lambda_\theta = \frac{r'd\theta'}{rd\theta} = \frac{r'}{r'-a}\frac{b-a}{b}, \quad (3)$$
$$\lambda_z = \frac{dz'}{dz} = 1.$$

At the inner boundary $r' = a$ of the cloak, it can be seen from Eq. (3) that $\lambda_\theta \to \infty$, while $\lambda_r$ and $\lambda_z$ are kept finite. Due to the infinite stretch $\lambda_\theta$ in the azimuthal direction, some components in mass density tensors and the modulus will approach infinite values near the inner boundary, as can be seen in the following

$$\rho'_r = \frac{\lambda_\theta\lambda_z}{\lambda_r} = \frac{r'}{r'-a},$$
$$\rho'_\theta = \frac{\lambda_r\lambda_z}{\lambda_\theta} = \frac{r'-a}{r'}, \quad (4)$$
$$\kappa' = \lambda_r\lambda_\theta\lambda_z = (\frac{b-a}{b})^2\frac{r'}{r'-a}.$$

The linear transformation of the spherical cloak is $r' = a + \frac{b-a}{b}r$, $\theta' = \theta$, $\phi' = \phi$, the corresponding principal stretches are

$$\lambda_r = \frac{dr'}{dr} = \frac{b-a}{b},$$
$$\lambda_\theta = \frac{r'd\theta'}{rd\theta} = \frac{r'}{r'-a}\frac{b-a}{b}, \quad (5)$$
$$\lambda_\phi = \frac{r'd\phi'}{rd\phi} = \frac{r'}{r'-a}\frac{b-a}{b}.$$

At the inner boundary $r' = a$, both $\lambda_\theta$ and $\lambda_\phi$ tend to infinite, according to Eq. (2), some components of mass density and the modulus will also become infinite, as explained in the following

$$\rho'_r = \frac{\lambda_\theta\lambda_\phi}{\lambda_r} = \frac{b-a}{b}(\frac{r'}{r'-a})^2,$$
$$\rho'_\theta = \frac{\lambda_r\lambda_\phi}{\lambda_\theta} = \frac{b-a}{b},$$
$$\rho'_\phi = \frac{\lambda_r\lambda_\phi}{\lambda_\theta} = \frac{b-a}{b} \quad (6)$$
$$\kappa' = \lambda_r\lambda_\theta\lambda_\phi = (\frac{b-a}{b})^3(\frac{r'}{r'-a})^2.$$

This 3D singularity does not exist in an electromagnetic cloak, which has the transformed material parameters as [13]

$$\boldsymbol{\varepsilon}' = \boldsymbol{\mu}' = \text{diag}[\frac{\lambda_1}{\lambda_2\lambda_3}, \frac{\lambda_2}{\lambda_3\lambda_1}, \frac{\lambda_3}{\lambda_1\lambda_2}]. \quad (7)$$

where two infinite principal stretches with the same order will never produce infinite material parameters. In order to remove the singularity of acoustic cloak, we should firstly use the other set of the transformed material parameters firstly obtained by Norris [9] that the mass density is isotropic and the modulus is anisotropic, i.e.,

$$\boldsymbol{\kappa}' = \text{diag}[\frac{\lambda_1}{\lambda_2\lambda_3}, \frac{\lambda_2}{\lambda_3\lambda_1}, \frac{\lambda_3}{\lambda_1\lambda_2}], \rho' = \frac{1}{\lambda_1\lambda_2\lambda_3}. \quad (8)$$

Obviously, the mass density has no singularity, and the modulus has the same form as that of electromagnetic material parameters. The non-unique transformation of acoustic material parameters and fields is also discussed in Ref. [14]. This kind of 3D acoustic cloak has no singularity, just as the 3D electromagnetic cloak, however the 2D acoustic cloak will have singular modulus at the inner boundary. We can use the similar method as the electromagnetic cloak [11] to remove the singularity of the 2D acoustic cloak. For the 2D cylindrical cloak, the stretch $\lambda_z$ perpendicular to the cloaking plane is decoupled with the in-plane stretches $\lambda_r$ and $\lambda_\theta$, so the continuously variant stretch $\lambda_z$ can be chosen freely except that $\lambda_z = 1$ must be satisfied at the outer boundary in order to have the fixed outer boundary. Thus, we can set $\lambda_z = \lambda_\theta$, the ratios of the stretches in Eq. (8) are finite and $\lambda_z = 1$ at the outer boundary. The resulting material parameters are obtained from Eq. (8) as

$$\kappa'_r = \frac{\lambda_r}{\lambda_\theta \lambda_z} = \frac{b}{b-a}(\frac{r'-a}{r'})^2,$$

$$\kappa'_\theta = \frac{\lambda_\theta}{\lambda_z \lambda_r} = \frac{b}{b-a}. \quad (9)$$

$$\rho' = \frac{1}{\lambda_r \lambda_\theta \lambda_z} = (\frac{b}{b-a})^3 (\frac{r'-a}{r'})^2.$$

This set of material parameters has no singularity, i.e., infinite value.

In order to validate the acoustic cloak given by Eq. (9), we use the PDE mode $\nabla \cdot (\mathbf{c} \nabla u) + au = 0$ (Helmholtz equation) of commercial software COMSOL Multiphysics to validate of the cloaking effect of the designed acoustic cloak. This equation has the same form of acoustic wave for scalar displacement $u$ [14]. Figure 1 shows the computational domain for a horizontally incident wave. To verify the effect of the cloak, in the PDE mode of Helmholtz equation, we let a plane acoustic wave be incident on this cloak, so the incident boundary of the computation domain is set to be Dirichlet condition $u = \exp(-ikx)$ where $k = 2\pi/\iota$ is the wave number and $\iota$ is the wavelength. The other boundaries of the computation domain are set to be Neumann boundary conditions $\mathbf{n} \cdot \nabla u = 0$, in order to avoid disturbance for the wave propagation. In the following simulation, the background medium is water $\rho = 1 \times 10^3$, $\kappa = 2.18 \times 10^9$ in SI units and the wavelength of the incident wave is $\iota = 0.35$ m. The material parameters within the cloaked area are $\rho_{inc} = \rho$ and $\kappa_{inc} = 1000\kappa$, respectively. Figure 2 shows the displacement field of the object with the designed cloak layer. It is clearly seen from the Fig. 2 that the designed cloak does not disturb the outside field and shield the obstacle inside of the cloak from detection for acoustic wave. The necessary material parameters for the cloak retrieved from the calculation are shown in figures 3(a), 3(b), 3(c), and 3(d) for $\kappa'_{xx}$, $\kappa'_{xy}$, $\kappa'_{yy}$ and the mass density $\rho'$, respectively. As shown in these figures, the cloaking effect can indeed be achieved by the finite material parameters.

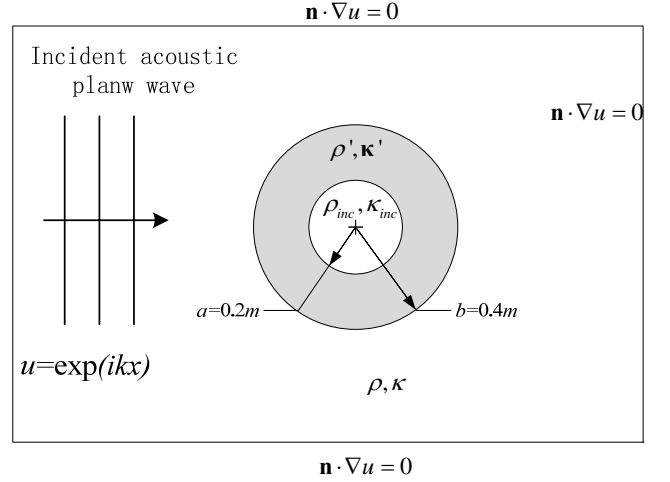

Figure 1. Computational domain and simulation for a left-incident cylindrical acoustic cloak based on Helmholtz equation.

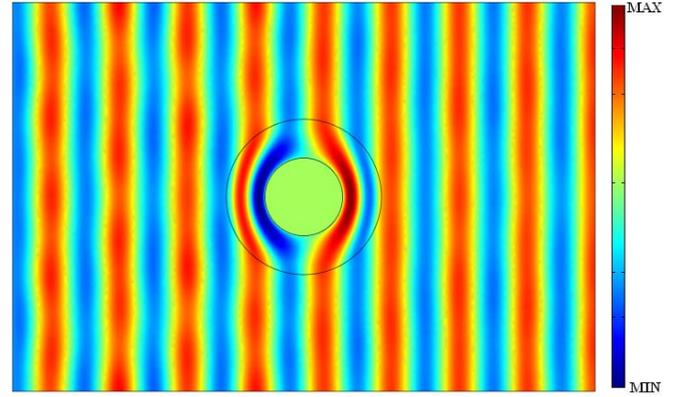

Figure 2. Simulation of the acoustic wave with the designed cloak layer

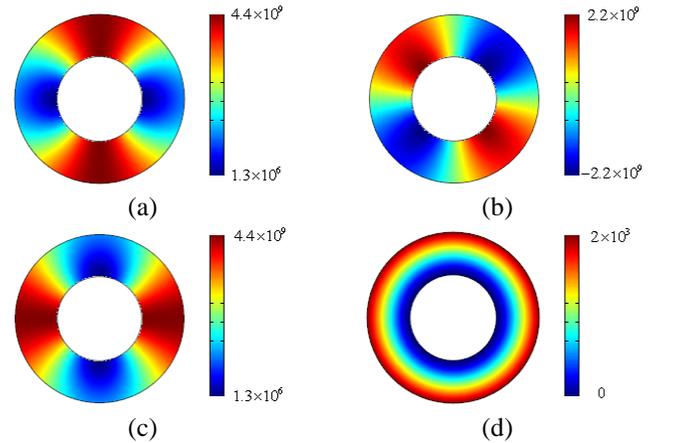

Figure 3. Values of the non-vanishing components of the bulk modulus tensors of the designed cloak, (a) $\kappa'_{xx}$, (b) $\kappa'_{xy}$, (c) $\kappa'_{yy}$, and (d) mass density $\rho'$.

## NONSINGULAR ARBITRARY SHAPED ACOUSTIC CLOAK

To generalize the above idea to 2D acoustic cloaks of arbitrary shape, we observe that the in-plane ($x_1$, $x_2$) and out-of-plane ($x_3$) deformations are decoupled, thus the out-of-plane stretch $\lambda_3$ is a principal one and can be chosen freely, with the restriction that it must equal to unity at the outer boundary. In order to make $\lambda_3$ tend to infinite with the same order as the in-plane infinite stretch near the inner boundary, we should firstly obtain the in-plane principal stretches. According to Hu et al [11], the in-plane deformation can be solved from the 2D Laplace's equation with proper boundary conditions

$$\frac{\partial^2 x_i}{\partial x'^2_1} + \frac{\partial^2 x_i}{\partial x'^2_2} = 0, \quad i=1,2,; \ (x'_1, x'_2) \in \Omega', \quad (10)$$
$$\mathbf{x}(\partial\Omega'_-) = 0, \ \mathbf{x}(\partial\Omega'_+) = \mathbf{x}'(\partial\Omega'_+).$$

where $\partial\Omega'_+$ and $\partial\Omega'_-$ are the outer and inner boundaries of the 2D cloak respectively, and the inner boundary is enlarged by the point located at (0,0). For the out-of-plane stretch, $x'_3 = x_3$ is assumed. The deformation gradient tensor $\mathbf{A} = \nabla\mathbf{x}'$ can be inversely obtained from Eq. (10), as well as the left Cauchy-Green deformation tensor $\mathbf{B} = \mathbf{A}\mathbf{A}^T$. In Cartesian coordinate system, we have

$$\mathbf{A} = \begin{bmatrix} A_{11} & A_{12} & 0 \\ A_{21} & A_{22} & 0 \\ 0 & 0 & 1 \end{bmatrix}, \ \mathbf{B} = \begin{bmatrix} B_{11} & B_{12} & 0 \\ B_{12} & B_{22} & 0 \\ 0 & 0 & 1 \end{bmatrix}. \quad (11)$$

The in-plane principal stretches can be calculated directly from $\mathbf{B}$ as

$$\lambda_{1,2} = \sqrt{\frac{B_{11} + B_{22} \mp \sqrt{B_{11}^2 - 2B_{11}B_{22} + B_{22}^2 + 4B_{12}^2}}{2}}. \quad (12)$$

$\lambda_2$ is always greater than $\lambda_1$ and it will tend to infinite near the inner boundary. To avoid the singularity, we let $\lambda_3$ tend to infinity with the same order as $\lambda_2$ at the inner boundary and $\lambda_3 = 1$ at the outer boundary. For an arbitrary 2D cloak, we cannot simply set $\lambda_3 = \lambda_2$, since there may be $\lambda_2 \neq 1$ at the outer boundary, which means the principle stretch may not be tangential to the outer boundary. For this reason, we choose the out-of-plane stretch $\lambda_3$ as

$$\tilde{\lambda}_3 = C_0 \left(\frac{|x'_1 - x_1| + |x'_2 - x_2|}{|x'_1| + |x'_2|}\right)\lambda_2 + 1. \quad (13)$$

where $C_0$ is a constant value that can be used to adjust the material parameters. Then Eq. (11) can be rewritten as

$$\tilde{\mathbf{A}} = \begin{bmatrix} A_{11} & A_{12} & 0 \\ A_{21} & A_{22} & 0 \\ 0 & 0 & \tilde{\lambda}_3 \end{bmatrix}, \ \tilde{\mathbf{B}} = \begin{bmatrix} B_{11} & B_{12} & 0 \\ B_{12} & B_{22} & 0 \\ 0 & 0 & \tilde{\lambda}_3^2 \end{bmatrix}. \quad (14)$$

Thus for an arbitrary 2D acoustic cloak, the nonsingular material parameters in the transformed space are finally given by Eq. (8), or

$$\kappa' = \tilde{\mathbf{B}}/\det\tilde{\mathbf{A}}, \quad (15)$$
$$\rho = 1/\det\tilde{\mathbf{A}}.$$

To illustrate the idea, an arbitrary cloak illuminated by a plane harmonic wave is examined, the computational domain is shown in Fig. 4, while the background medium and material parameters within the cloaked area are as same as the above cylindrical cloak, and the wavelength is also 0.35m. Fig. 5 shows the left incident acoustic wave without the cloak layer. The strong scattered displacement fields can be observed. The simulation results of an arbitrary shaped 2D cloak designed by the proposed method are given in Fig. 6, where $C_0 = 1$ is taken in Eq. (13). The corresponding material parameters are shown in Fig. 7. It can be seen that the cloaking effect can really be achieved by finite material parameters.

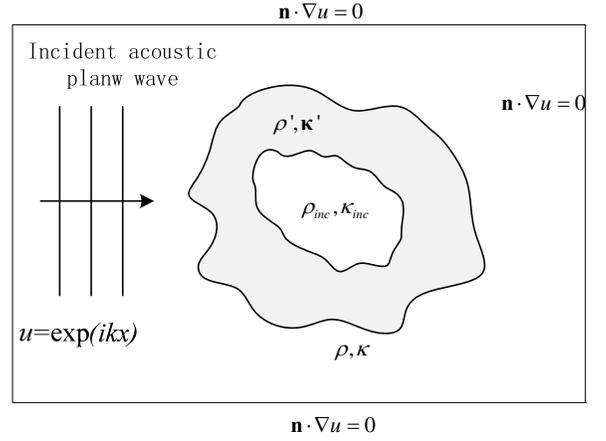

Figure 4. Computational domain and simulation for a left-incident arbitrary shaped acoustic cloak based on Helmholtz equation.

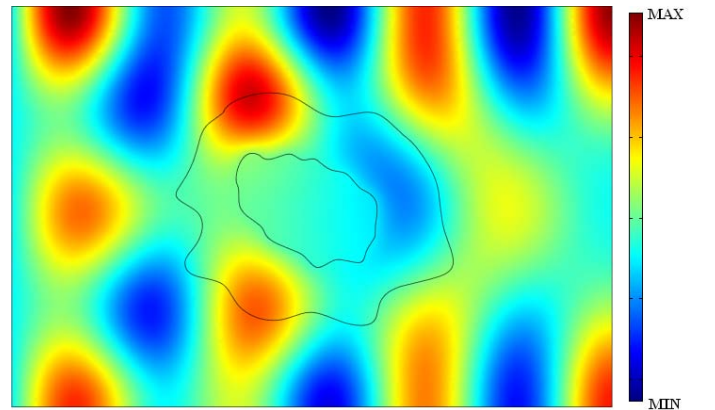

Figure 5. Simulation of a left incident acoustic wave without the cloak layer.

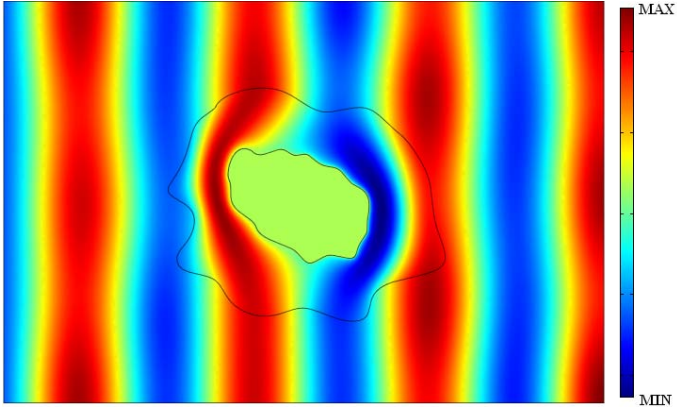

Figure 6. Simulation of a left incident acoustic wave with the designed cloak layer.

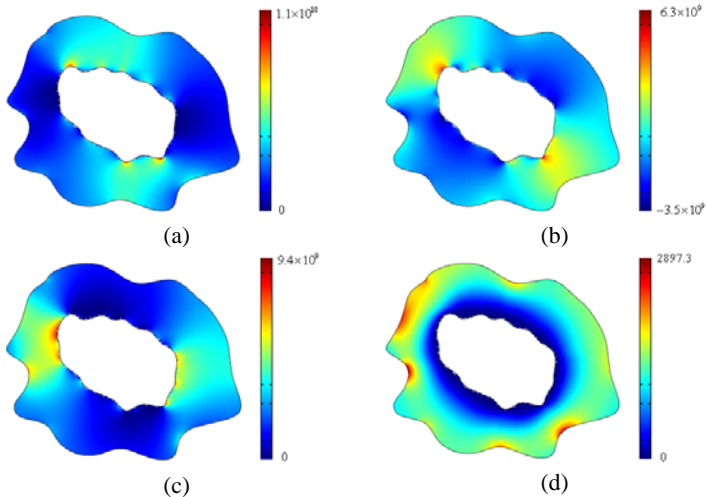

Figure 7. Values of the non-vanishing components of the bulk modulus mass density tensors of the designed cloak, (a) $\kappa'_{xx}$, (b) $\kappa'_{xy}$, (c) $\kappa'_{yy}$, and (d) mass density $\rho'$.

**DISCUSSION AND CONCLUSION**

With help of deformation view on the transformation method, we show that by adjusting the principle stretch out of the cloaking plane, 2D acoustic cloaks of arbitrary shapes without singularity can be constructed. Since the conventional isotropic-modulus acoustic cloaks will have material singularity, thus, we suggest to construct nonsingular acoustic cloak with isotropic-inertia transformation material parameters, which is originally proposed by Norris [9]. For this type of acoustic cloak, the expressions for the transformed material parameters are the same as the counterpart of the electromagnetic case, so the method proposed for a nonsingular 2D electromagnetic cloak can be applied directly.

It is also worth to mention that the proposed method cannot remove the zero-value material parameters in the transformation media, which may be tackled with help of non-Euclidean cloaking theory [15].

In conclusion, we have proposed a method to design arbitrary shaped 2D isotropic-inertia acoustic cloaks without singularity. Examples of cylindrical acoustic cloak and arbitrary shaped acoustic cloak are given. The proposed method is based on the spatial transformation without any other additional assumptions, it can be directly extended to other problem, such as thermal conduction designed based on coordinate transformations method.

**ACKNOWLEDGMENTS**

This work is supported by the National Natural Science Foundation of China (90605001, 10702006, 10832002), the National Basic Research Program of China (2006CB601204), and Beijing Municipal Commission of Education Project (20080739027).